\documentclass[a4paper,11pt]{article}
\usepackage{braket}
\usepackage{pos_1
}

\title{Neutrino oscillations in astrophysical environment accounting for neutrino charge radius and anapole moment}
\ShortTitle{Neutrino oscillations accounting for neutrino charge radius and anapole moment}

\author*[a]{Vadim Shakhov}
\author[a]{Alexander Studenikin}

\affiliation[a]{Department of Theoretical Physics, Lomonosov Moscow State University,\\
  119992 Moscow, Russia}

\emailAdd{schakhov.vv15@physics.msu.ru}
\emailAdd{studenik@srd.sinp.msu.ru}
\emailAdd{a-studenik@yandex.ru}

\abstract{We derive an effective neutrino evolution Hamiltonian and corresponding expression for neutrino spin oscillations probabilities accounting for neutrino
interactions with external electric current due to neutrino charge radii and anapole moment. The
results are interesting for possible applications in astrophysics.}

\FullConference{%
  41st International Conference on High Energy physics - ICHEP2022\\
  6-13 July, 2022\\
  Bologna, Italy
}

\begin{document}
\maketitle

Neutrino flavour, spin and spin-flavour oscillations engendered by neutrino interactions with an external electric current due to neutrino charge radii and anapole moments are investigated. We consider two flavour neutrinos with two possible helicities $\nu_{f}=\left(\nu_{e}^{-}, \nu_{x}^{-}, \nu_{e}^{+}, \nu_{x}^{+}\right)$ and perform calculations that are analogous to those in \cite{Fabbricatore:2016nec, Pustoshny:2018jxb, Abdullaeva:2021zxj}. In the mass basis the neutrino effective potential  describing electromagnetic interactions of the  neutrino field $\nu$ with the external electric current is given by
\begin{equation}
    H_J^{(m)fi} = \lim_{q \rightarrow 0}\frac1T\frac{\left<\nu_f(p_f,h_f)\right|\int d^4x \mathcal{H}_{J}\left|\nu_i(p_i,h_i)\right>}{\left<\nu(p,h)|\nu(p,h)\right>},
\end{equation}
where $q=p_i-p_f$, $p_i$ and $p_f$ are the initial and final neutrino momenta, $T$ is the normalization time.
The matrix element
\begin{equation}
    \left<\nu_f(p_f,h_f)\right|\mathcal{H}_{J}\left|\nu_i(p_i,h_i)\right> = \overline{u}_f(p_f, h_f)\Lambda_{\mu}^{f i}(q)\frac{1}{q^2}u_i(p_i, h_i)J^{\mu}_{EM}e^{-iqx}
\end{equation}
is dertermined by the electric current of the external charged fermions $f$ (the protons or electrons), $J^{\mu}_{EM} = e(n_f, n_f\mathbf v_f)$. Note that  $\Lambda_\mu^{fi}$ in (2) contains the corresponding terms of the neutrino electromagnetic vertex (for its decomposition see \cite{Giunti:2014ixa}). We are interested only in the charge radius $\braket{r^2}$ and anapole moment $f_A$, therefore, the electromagnetic vertex reads

\begin{equation}
\Lambda_\mu^{fi} (q) = (q^2 \gamma_\mu - q_\mu \gamma_\nu q^\nu) \left[\dfrac{\braket{r^2}^{fi}}{6} + f_A^{fi} \gamma_5 \right]
.
\end{equation}
This vertex in the flavour basis $\nu_f = (\nu^L_e, \nu^L_x, \nu^R_e, \nu^R_x)$ gives the following effective interaction Hamiltonian
\begin{equation}
H_J^{f} = H_{J_{\|}}^{(f)} + H_{J_{\perp}}^{(f)},
\end{equation}
where the parts of the interaction  Hamiltonian corresponding to the longitudinal $\boldsymbol{J}^{EM}_{\|}$ (with respect to $z$-axis of the neutrino propagation) and the transversal $\boldsymbol{J}^{EM}_{\perp}$ components of the electric current are given by

\begin{equation}\label{HJlong}\small
H_{J_{\|}}^{(f)}=2J^{EM}_{\|}
\begin{pmatrix}
\frac{\langle r^2 \rangle^{ee}}{6} - f_A^{ee} & \ \ \ \ \ \frac{\langle r^2 \rangle^{ex}}{6} - f_A^{ex} & 0 & 0 \\
\frac{\langle r^2 \rangle^{ex}}{6} - f_A^{ex} & \ \ \ \ \ \frac{\langle r^2 \rangle^{xx}}{6} - f_A^{xx} & 0 & 0 \\
0 & 0 & \frac{\langle r^2 \rangle^{ee}}{6} + f_A^{ee} & \ \ \ \ \ \frac{\langle r^2 \rangle^{ex}}{6} + f_A^{ex} \\
0 & 0 & \frac{\langle r^2 \rangle^{ex}}{6} + f_A^{ex} & \ \ \ \ \ \frac{\langle r^2 \rangle^{xx}}{6} + f_A^{xx}
\end{pmatrix},
\end{equation}

\begin{equation}\label{HJperp}
\footnotesize
H_{J_{\perp}}^{(f)}=2J^{EM}_{\perp}
\begin{pmatrix}
0 & 0 & \left(\frac{f_A}{\gamma}\right)_{ee}e^{i\chi}  & H^{+}e^{i\chi}  \\
0 & 0 & H^{-}e^{i\chi} & \left(\frac{f_A}{\gamma}\right)_{xx}e^{i\chi} \\
\left(\frac{f_A}{\gamma}\right)_{ee}e^{-i\chi} & H^{-}e^{-i\chi}  & 0 & 0 \\
H^{+}e^{-i\chi} & \left(\frac{f_A}{\gamma}\right)_{xx}e^{-i\chi} & 0 & 0
\end{pmatrix}
,
\end{equation}
where $\chi$ is the angle between the fixed $x$-axis that is perpendicular to $z$-axis, $H^{\pm} = \left(\frac{f_A}{\gamma}\right)_{ex} \pm \tilde{\gamma}_{12}^{-1} \frac{\braket{r^2}^{12}}{6}$ and

\begin{equation}\label{GammaFactor}
\gamma_{\alpha}^{-1} =\frac{m_{\alpha}}{E_{\alpha}} , \ \ \
\gamma_{\alpha\beta}^{-1} =\frac12\left(\gamma_{\alpha}^{-1} + \gamma_{\beta}^{-1}\right) , \ \ \
\tilde \gamma_{\alpha\beta}^{-1} =\frac12\left(\gamma_{\alpha}^{-1} - \gamma_{\beta}^{-1}\right) ,
\end{equation}

\begin{equation}
\small
\begin{aligned}
&\langle r^2 \rangle^{ee} = \langle r^2 \rangle^{11}\cos^2\theta + \langle r^2 \rangle^{22} \sin^2\theta + \langle r^2 \rangle^{12} \sin2\theta,
&f_A^{ee} &= f_A^{11}\cos^2\theta + f_A^{22} \sin^2\theta + f^{12}_A \sin2\theta,\\
&\langle r^2 \rangle^{xx} = \langle r^2 \rangle^{11}\sin^2\theta + \langle r^2 \rangle^{22}\cos^2\theta - \langle r^2 \rangle^{12}\sin2\theta,
&f_A^{xx} &= f_A^{11}\sin^2\theta + f_A^{22}\cos^2\theta - f_A^{12}\sin2\theta,\\
&\langle r^2 \rangle^{ex} = \langle r^2 \rangle^{12} \cos2\theta + \dfrac12 \left( \langle r^2 \rangle^{22} - \langle r^2 \rangle^{11} \right)\sin2\theta,
&f_A^{ex} &= f_A^{12} \cos2\theta + \dfrac12 \left( f_A^{22} - f_A^{11} \right)\sin2\theta,
\end{aligned}
\end{equation}
\begin{equation}
\small
\begin{aligned}
\left(\dfrac{f_A}{\gamma}\right)_{ee} = \dfrac{f_A^{11}}{\gamma_{11}} \cos^2\theta + &\dfrac{f_A^{22}}{\gamma_{22}} \sin^2 \theta + \dfrac{f_A^{12}}{\gamma_{12}} \sin 2 \theta , \ \
\left(\dfrac{f_A}{\gamma}\right)_{xx} = \dfrac{f_A^{11}}{\gamma_{11}} \sin^2\theta + \dfrac{f_A^{22}}{\gamma_{22}} \cos^2 \theta - \dfrac{f_A^{12}}{\gamma_{12}} \sin 2 \theta , \\
&\left(\dfrac{f_A}{\gamma}\right)_{ex} =\dfrac{f_A^{12}}{\gamma_{12}} \cos 2 \theta + \dfrac 1 2 \left( \dfrac{f_A^{22}}{\gamma_{22}} - \dfrac{f_A^{11}}{\gamma_{11}} \right) \sin 2 \theta.
\end{aligned}
\end{equation}

It can be seen that the neutrino charge radius and anapole moment interactions with the longitudinal current  $\boldsymbol{J}^{EM}_{\|}$ modifies the neutrino flavour oscillations, and the neutrino interactions with the transversal electric current $\boldsymbol{J}^{EM}_{\perp}$ can generate neutrino spin and spin-flavour oscillations.

Neutrinos can also interact with moving matter and  with an external magnetic field through neutrino magnetic moment. For interactions with moving matter one gets (see \cite{Pustoshny:2018jxb})
\begin{equation}\footnotesize
H_{mat}= \frac{G_F}{2\sqrt{2}}
\begin{pmatrix}
2(2 n_e -  n_n)\left(1-v_{||}\right)  & 0 & (2 n_e -  n_n)  v_{\perp}\left(\frac{\eta}{\gamma}\right)_{ee}  & (2 n_e -  n_n) v_{\perp} \left(\frac{\eta}{\gamma}\right)_{ex} \\

0 & - 2n_n\left(1-v_{||}\right)  &  - n_n v_{\perp} \left(\frac{\eta}{\gamma}\right)_{ex} & - n_n v_{\perp} \left(\frac{\eta}{\gamma}\right)_{xx}  \\

(2 n_e -  n_n) v_{\perp} \left(\frac{\eta}{\gamma}\right)_{ee}  & - n_n v_{\perp} \left(\frac{\eta}{\gamma}\right)_{ex} & 0 & 0 \\

(2 n_e -  n_n) v_{\perp} \left(\frac{\eta}{\gamma}\right)_{ex} & - n_n v_{\perp}  \left(\frac{\eta}{\gamma}\right)_{xx} & 0 & 0
\end{pmatrix}
,
\end{equation}
where $n_n$ and $n_e$ are the neutron and electron number densities, $\bf{v} = \bf{v}_{||} + \bf{v}_{\perp}$ is the matter velocity and
\begin{equation}\footnotesize
\left(\frac{\eta}{\gamma}\right)_{ee} = \frac{\cos^2\theta}{\gamma_{11}} + \frac{\sin^2\theta}{\gamma_{22}} ,\ \ \ \ \
\left(\frac{\eta}{\gamma}\right)_{xx} = \frac{\sin^2\theta}{\gamma_{11}} + \frac{\cos^2\theta}{\gamma_{22}} ,\ \ \ \ \
\left(\frac{\eta}{\gamma}\right)_{ex} = \frac{\sin2\theta}{\tilde \gamma_{21}}
.
\end{equation}

The Hamiltonian that accounts for the neutrino magnetic moment interaction with two components of an external magnetic field $B = B_{||} + B_{\perp}$ reads
\begin{equation}\footnotesize
H_B=
\begin{pmatrix}
\left(\frac{\mu}{\gamma}\right)_{ee}B_{||} & \left(\frac{\mu}{\gamma}\right)_{ex}B_{||} & -\mu_{ee}B_{\perp}e^{i\phi} & -\mu_{ex}B_{\perp}e^{i\phi} \\

\left(\frac{\mu}{\gamma}\right)_{ex}B_{||} & \left(\frac{\mu}{\gamma}\right)_{xx}B_{||} & -\mu_{ex}B_{\perp}e^{i\phi} & -\mu_{xx}B_{\perp}e^{i\phi} \\

-\mu_{ee}B_{\perp}e^{-i\phi} & -\mu_{ex}B_{\perp}e^{-i\phi} & -\left(\frac{\mu}{\gamma}\right)_{ee}B_{||} & -\left(\frac{\mu}{\gamma}\right)_{ex}B_{||} \\

-\mu_{ex}B_{\perp}e^{-i\phi} & -\mu_{xx}B_{\perp}e^{-i\phi} & -\left(\frac{\mu}{\gamma}\right)_{ex}B_{||} & -\left(\frac{\mu}{\gamma}\right)_{xx}B_{||}
\end{pmatrix}
,
\end{equation}
where $\phi$ is the angle between $\textbf{v}_{\perp}$ and $\textbf{B}_{\perp}$. The components of the effective magnetic moment in the flavour basis $\mu_{\alpha \beta}$ are expressed through the components $\mu_{ij}$ in the mass basis,
\begin{equation}
\begin{aligned}\small
&\mu_{ee} = \mu_{11}\cos^2\theta + \mu_{22}\sin^2\theta + \mu_{12}\sin 2\theta ,
&\left(\frac{\mu}{\gamma}\right)_{ee} &= \frac{\mu_{11}}{\gamma_{11}}\cos^2\theta + \frac{\mu_{22}}{\gamma_{22}}\sin^2\theta + \frac{\mu_{12}}{\gamma_{12}}\sin 2\theta ,\\
&\mu_{xx} = \mu_{11}\sin^2\theta + \mu_{22}\cos^2\theta - \mu_{12}\sin 2\theta ,
&\left(\frac{\mu}{\gamma}\right)_{xx} &= \frac{\mu_{11}}{\gamma_{11}}\sin^2\theta + \frac{\mu_{22}}{\gamma_{22}}\cos^2\theta - \frac{\mu_{12}}{\gamma_{12}}\sin 2\theta,\\
&\mu_{ex} = \mu_{12}\cos2\theta + \frac12\left(\mu_{22} - \mu_{11}\right)\sin 2\theta ,
&\left(\frac{\mu}{\gamma}\right)_{ex} &= \frac{\mu_{12}}{\gamma_{12}}\cos2\theta + \frac12\left(\frac{\mu_{22}}{\gamma_{22}} - \frac{\mu_{11}}{\gamma_{11}}\right)\sin 2\theta.\\
\end{aligned}
\end{equation}
While deriving the neutrino oscillations probabilities we take into account all interaction Hamiltonians and also the vacuum Hamiltonian.

Consider two neutrino states with different helicities: $(\nu^L_e,\nu^R_{e})$. For the oscillation $\nu_{e}^{L} \leftrightarrow \nu_{e}^{R}$ probability we get

\begin{equation} \label{prob}
\begin{array}{c} {P_{\nu_{e}^{L} \rightarrow \nu_{e}^{R}}(x) =\sin^22\theta_{\mathrm{eff}}\sin ^{2} \frac{\pi x}{L_{\mathrm{eff}}},\ \ \
\sin^22\theta_{\mathrm{eff}} = \frac{E_{\mathrm{eff}}^{2}}{E_{\mathrm{eff}}^{2}+\Delta_{\mathrm{eff}}^{2}}},\ \ \
{L_{\mathrm{eff}} =\frac{\pi}{\sqrt{E_{\mathrm{eff}}^{2}+\Delta_{\mathrm{eff}}^{2}}}}, \end{array}
\end{equation}
where $E_{eff}^2$ and $\Delta _{eff}^2$ are expressed in terms of the elements of the Hamiltonian:
\small\begin{multline}
E_{eff}^2 = 4\left[\frac{G_F}{2\sqrt{2}} (2 n_e -  n_n) v_{\perp}\left(\frac{\eta}{\gamma}\right)_{ee} - \mu_{ee}B_{\perp}\cos \phi +\right.\\
\left. + 2J^{EM}_{\perp} \left(\frac{f_A}{\gamma}\right)_{ee} \cos \chi \right]^2 + 4\left[\mu_{ee}B_{\perp}\sin \phi - 2J^{EM}_{\perp}\left(\frac{f_A}{\gamma}\right)_{ee} \sin \chi\right]^2,
\end{multline}
\begin{equation}\small
\Delta _{eff}^2 = \left[ \frac{G_F}{\sqrt{2}}(2 n_e -  n_n)\left(1-v_{||}\right) + 2 \left(\frac{\mu}{\gamma}\right)_{ee}B_{||} - 4J^{EM}_{\|} f_A^{ee}    \right]^2
.
\end{equation}
\normalsize
It follows that whereas the spin oscillations can be generated by the neutrino anapole moment interactions with an external electric current, the interaction due to the  charge radius does not produce the spin oscillations. Thus, these peculiarities can be used for disintegration of the anapole moment and charge radius effects in neutrino interactions.

Now consider two neutrino flavour states with two different helicities: $(\nu^L_e,\nu^R_{x})$. For the case of the neutrino oscillations $\nu_{e}^{L} \leftrightarrow \nu_{x}^{R}$ the probability is again given by (\ref{prob}) with
\small
\begin{multline}
E_{eff}^2 = 4 \left[\frac{G_F}{2\sqrt{2}} (2 n_e -  n_n)  v_{\perp}\left(\frac{\eta}{\gamma}\right)_{ex} - \mu_{ex}B_{\perp}\cos \phi +
2J^{EM}_{\perp} \left[\tilde{\gamma}_{12}^{-1} \frac{\braket{r^2}^{12}}{6} + \left(\frac{f_A}{\gamma}\right)_{ex}\right]\cos \chi\right]^2 +\\
+4 \left[\mu_{ex}B_{\perp}\sin \phi -
J^{EM}_{\perp} \left[\tilde{\gamma}_{12}^{-1} \frac{\braket{r^2}^{12}}{6} + \left(\frac{f_A}{\gamma}\right)_{ex}\right]\sin \chi \right]^2
,
\end{multline}

\begin{multline}
\Delta _{eff}^2 = \left[-\dfrac{\Delta m^2}{2 E}\cos \theta + \frac{G_F}{\sqrt{2}}(2 n_e -  n_n)\left(1-v_{||}\right)
+\left[\left(\frac{\mu}{\gamma}\right)_{ee} + \left(\frac{\mu}{\gamma}\right)_{xx}\right]B_{||} +\right.\\
\left. +J^{EM}_{\|} \left(\frac{\langle r^2 \rangle^{ee} - \langle r^2 \rangle^{xx}}{6} - f_A^{ee} + f_A^{xx}\right) \right]^2
.
\end{multline}
\normalsize
Contrary to the previous case, these expressions depend on both the neutrino anapole moment and the charge radius.

The work is supported by the Russian Science Foundation under grant No.22-22-00384.
V.S. acknowledges the support from the National Center for Physics and Mathematics (Project “Study of coherent elastic neutrino-atom and -nucleus scattering and neutrino electromagnetic properties using a high-intensity tritium neutrino source”).


\begin{thebibliography}{99}
%
\bibitem{Fabbricatore:2016nec}
R.~Fabbricatore, A.~Grigoriev and A.~Studenikin,
{\it J. Phys. Conf. Ser.} {\bf{718}} (2016) 062058.

\bibitem{Pustoshny:2018jxb}
P.~Pustoshny and A.~Studenikin,
{\it Phys. Rev. D } {\bf{98}} (2018) 113009.

\bibitem{Abdullaeva:2021zxj}
U.~Abdullaeva, V.~Shakhov, A.~Studenikin and A.~Tsvirov,
{\it J. Phys. Conf. Ser.} {\bf{2156}} (2021) 012229.

\bibitem{Giunti:2014ixa}
C.~Giunti and A.~Studenikin,
{\it Rev. Mod. Phys.} {\bf{87}} (2015) 531.

\end{thebibliography}
\end{document}